# Skyrmion phase and competing magnetic orders on a breathing kagomé lattice

Max Hirschberger[1]*, Taro Nakajima[1]*, Shang Gao[1], Licong Peng[1], Akiko Kikkawa[1], Takashi Kurumaji[1‡], Markus Kriener[1], Yuichi Yamasaki[2,3], Hajime Sagayama[4], Hironori Nakao[4], Kazuki Ohishi[5], Kazuhisa Kakurai[1,5], Yasujiro Taguchi[1], Xiuzhen Yu[1], Takahisa Arima[1,6], and Yoshinori Tokura[1,7]

[1]RIKEN Center for Emergent Matter Science (CEMS), Wako 351-0198, Japan.

[2]Research and Services Division of Materials Data and Integrated System (MaDIS), National Institute for Materials Science (NIMS), Tsukuba 305-0047, Japan

[3]PRESTO, Japan Science and Technology Agency (JST), Kawaguchi 332-0012, Japan

[4]Institute of Materials Structure Science, High Energy Accelerator Research Organization, Tsukuba, Ibaraki 305-0801, Japan

[5]Neutron Science and Technology Center, Comprehensive Research Organization for Science and Society (CROSS), Tokai, Ibaraki 319-1106, Japan

[6]Department of Advanced Materials Science, University of Tokyo, Kashiwa, Chiba 277-8561, Japan

[7]Department of Applied Physics and Tokyo College, University of Tokyo, Bunkyo-ku 113-8656, Japan

*: equal contribution

‡: Current address: Department of Physics, Massachusetts Institute of Technology, Cambridge, Massachusetts 02139, USA

## Abstract.

**Magnetic skyrmion textures are realized mainly in non-centrosymmetric, e.g. chiral or polar, magnets. Extending the field to centrosymmetric bulk materials is a rewarding challenge, where the released helicity / vorticity degree of freedom and higher skyrmion density result in intriguing new properties and enhanced functionality. We report here on the experimental observation of a skyrmion lattice (SkL) phase with large topological Hall effect and an incommensurate helical pitch as small as 2.8 nm in metallic $Gd_3Ru_4Al_{12}$, which materializes a breathing kagomé lattice of Gadolinium moments. The magnetic structure of several ordered phases, including the SkL, is determined by resonant x-ray diffraction as well as small angle neutron scattering. The SkL and helical phases are also observed directly using Lorentz transmission electron microscopy. Among several competing phases, the SkL is promoted over a low-temperature transverse conical state by thermal fluctuations in an intermediate range of magnetic fields.**

# Introduction.

The magnetic skyrmion lattice (SkL) is a periodic array of spin vortices, which may be considered as an assembly of individual, tubular skyrmion particles protected against decay by their topological winding number [1-3]. Skyrmions hold significant technological promise, for example as tiny bits for data storage devices [4], which are highly controllable by small applied electrical currents [5]. From a fundamental perspective, we may classify nearly all previously reported real-world realizations of the SkL state into two families: (1) Thin ferromagnetic slabs, where dipolar interactions enable formation of topological bubbles with characteristic size of $\lambda$=100 nm-10 μm [6]. (2) Magnets with broken inversion symmetry (i.e. chiral or polar structures) [1-3,7], where competing Heisenberg and Dzyaloshinskii-Moriya interactions favor twisted spin structures with $\lambda$~10-200 nm. Magnetic interfaces, inversion-breaking by definition, are included in this second category [8,9].

In the search for even smaller skyrmions ($\lambda$=1-10 nm), which are expected to show giant responses to optical, electrical, and magnetic stimuli [4], attention has turned to systems with competing exchange interactions, or generalized Ruderman-Kittel-Kasuya-Yosida (RKKY) interactions [10-15]. Noteworthy in this context is the paradigm shift concerning the role of spin-orbit coupling: Centrosymmetric materials, where Dzyaloshinskii-Moriya interactions are absent or cancel out globally, have now moved into the cross-hairs of the search for skyrmion host compounds. They offer a path not only towards the miniaturization of spin textures, but also towards new physical properties related to the energetic near-degeneracy of left- and right-handed screws as well as Néel and Bloch type spin helicities [10, 11, 16]. Amongst these predicted properties are time-dependent helicity changes in response to an applied current [11], a modified excitation spectrum [11], mixed phases and near-degeneracy of skyrmions and antiskyrmions [10], background-free generation of second harmonic light due to a large toroidal moment [17], and qualitatively new decay mechanisms [18].

The recently discovered SkL phase with giant topological Hall effect in centrosymmetric $Gd_2PdSi_3$, with a triangular lattice of classical $Gd^{3+}$ spins, represents a seminal step into the direction of realizing the theoretical predictions listed above [19]. However, $Gd_2PdSi_3$ suffers from unavoidable disorder in the (Pd,Si) sublattice and a related crystallographic superstructure, rendering it challenging to perform advanced real-space imaging techniques [19].

In our present work, we reveal the SkL and competing magnetic orders in centrosymmetric $Gd_3Ru_4Al_{12}$, a good metal ($\rho_{300K}/\rho_{2K}$~6) in which the number of magnetic moments per layer and skyrmion is even smaller than in $Gd_2PdSi_3$. Using a combination of real-space imaging, x-ray and neutron scattering, as well as measurements of the topological Hall effect, the presence of the SkL is established unambiguously. On the basis of our experimental observations, we discuss how a combination of frustrated RKKY interactions, local ion anisotropy, and thermal fluctuations in $Gd_3Ru_4Al_{12}$ provides the first opportunity to observe metastable skyrmions in a centrosymmetric material.

# Results.

**Structural properties.** $Gd_3Ru_4Al_{12}$ crystallizes in hexagonal space group $P6_3/mmc$ with weak, yet finite, anisotropy in the magnetization and transport properties [20-23]. We label unit vectors aligned with the principal crystallographic axes as **a**, **b**, and **c** (Fig. 1**a,b**). For illustrative purposes, the structure may be decomposed into buckled $Ru_4Al_8$ layers (containing the inversion center) and perfectly planar $Gd_3Al_4$ sheets stacked along the *c*-axis (Fig. 1**a**). The magnetic moments at the rare earth (Gd) site materialize a breathing kagomé network (Fig. 1**b**), equivalent to a triangular lattice of trimer plaquettes formed by sets of three Gd moments [22, 23]. A quantitative measure of the distortion away from the ideal Kagome structure is the ratio between alternating (breathing) nearest ($r$) and next nearest ($r'$) neighbor Gd-Gd distances, $r/r'=0.73$ (Fig. 1**b**). Information about crystal growth and characterization, as well as a discussion of standard experimental techniques used in this work, is provided in the Methods section and in Supplementary Note 1.

**Magnetic phase diagram from magnetization and transport experiments.** In good agreement with previous reports [22,23], we observe Curie-Weiss behavior in the magnetic susceptibility $M/H$ at temperature $T > 200$ K (Fig. 1**c**). The dominant ferromagnetic interaction, associated with magnetic coupling within the trimer plaquette, is evident in the high Curie-Weiss temperature $T_{CW} \approx 70$ K. Long-range magnetic order sets in far below $T_{CW}$ at $T_{N2} = 18.6$ K, underlining the importance of competing magnetic interactions in this material, likely RKKY couplings which oscillate as a function of Gd-Gd distance [22]. Two sharp anomalies in $c_P(T)$ at $T_{N1}$ and $T_{N2}$ (Fig. 1**d**), a double kink profile of the magnetic susceptibility *M/H* (Fig. 1**e**), as well as two changes of slope in the zero-field longitudinal resistivity $\rho_{xx}$(T) (Supplementary Figure 5) all suggest successive evolution of order parameters, as is frequently the case in magnets with competing interactions [24-26]. Microscopic evidence for the phase transitions in zero field was obtained using elastic x-ray scattering (REXS), in resonance with the Gadolinium $L_2$ absorption edge and with polarization analysis at the detector. Sinusoidal magnetic order in the hexagonal *a-b* plane at $T_{N2}$ = 18.6 K can be separated from the onset of three-dimensional helical order at $T_{N1}$ = 17.2 K (Fig. 1**f**). As the incommensurate magnetic modulation vector **q** was found to be aligned within the hexagonal plane in the scattering experiments (Fig. 1**g**), six directions of **q** are equivalent by symmetry. We label three of these directions by $\mathbf{q}_1$, $\mathbf{q}_2$, and $\mathbf{q}_3$ (inset of Fig. 1**c**). The $\mathbf{q}_i$ vectors are locked to the **a*** (and equivalent) directions, where **a*** and **b*** are unit vectors in reciprocal space. The helical pitch length corresponding to **q** is $\lambda$ = 2.8 nm at *T* = 2.4 K, much smaller than in typical non-centrosymmetric B20 compounds such as MnSi ($\lambda \approx$ 19 nm) [2].

In the following, we establish the magnetic phase diagram using comprehensive measurements of the magnetic susceptibility $\chi_{DC}=\partial M/\partial H$ and Hall conductivity $\sigma_{xy}$. We have corrected the significant demagnetization effect for these data sets, and more generally for all data recorded under isothermal conditions (as well as all phase diagrams). In this spirit, isothermal data are plotted as a function of the internal magnetic field $H_{int}$ = *H-NM*, where *H* is the externally applied magnetic field, *N* is the demagnetization factor, and *M* is the bulk magnetization (Methods and Supplementary Table 2).

In the configuration $H \mathbin{/\mkern-3mu/} c$, the degeneracy of the $\mathbf{q}_i$ is maintained. Several magnetic phase boundaries are marked by open black symbols in the contour plot of $\chi_{DC}$ (Fig. 2**a**). In anticipation of the REXS results of Fig. 3, we label this cornucopia of competing magnetic states as helical (H), transverse conical (TC), fan-like (F), skyrmion lattice (SkL), field-aligned ferromagnet (FA-FM), and the as-yet unidentified phase V. Raw data of $\chi_{DC}$ are presented in Supplementary Figure 7.

Out of this large number of magnetic phases, the SkL is distinguished by a large topological Hall effect (THE) due to the non-zero integer winding number of the magnetic texture and the resulting Berry curvature of conduction electrons [27,28]. This transport signature provides direct evidence for the chiral nature of the magnetic order in the SkL phase. In Fig. 2**b**, a box-shaped and strongly field- and temperature-hysteretic Hall conductivity signal (shaded in grey) emerges on the back of a smooth background in an intermediate range of magnetic fields. We approximate the background by a low-order (odd) polynomial and extract the topological Hall conductivity $\sigma_{xy}^{THE}$. The topological signal as obtained from the isothermal field scans is confined within the boundaries of the SkL phase (Fig. 2**c**). Meanwhile, measurements of $\sigma_{xy}(T)$ at fixed magnetic field and for increasing temperature ($dT/dt > 0$) show a large split between curves recorded under zero-field cooled (ZFC) and field-cooled (FC) sample conditions, exclusively at intermediate field values (Fig. 2**d**, $\mu_0 H_{int} = 1.22$ and $1.42$ T). The natural conclusion is that a metastable SkL state, with its largely enhanced $\sigma_{xy}(T)$, can be sustained at the lowest temperatures in the FC experiment, where the SkL is absent under ZFC conditions. This behavior suggests the stabilization of the SkL by thermal fluctuations (c.f. Discussion section). The point of divergence between the ZFC and FC curves at $T$=5-8 K in Fig. 2**d** marks the first order phase transition between the TC and SkL states in our phase diagram (labeled in Fig. 2**a,c** by black open squares). Detailed susceptibiltity measurements evidence that the boundaries of phases TC and SkL with all surrounding phases are also strongly of first order (Supplementary Figure 7).

**Resonant elastic x-ray scattering (REXS) and microscopic magnetic structure.** We now proceed to study the field-induced magnetic phases using REXS (Fig. 3) and real-space imaging (Fig. 4, next section), before finally returning to a semiquantitative analysis of the Hall signal. For polarization analysis in REXS, three mutually orthogonal components of the **q**-modulated magnetic moment **m(q)** are separated viz. [34]

$$\mathbf{m}(\mathbf{q}) = \mathbf{c}\, m_{//c}(\mathbf{q}) + \hat{\mathbf{q}}\, m_{//q}(\mathbf{q}) + (\mathbf{c} \times \hat{\mathbf{q}})\, m_{\perp q,c}(\mathbf{q}) \qquad (1)$$

where $\hat{\mathbf{q}}$ is a vector of unit length along **q**. In our experiment, the incoming beam of x-rays is linearly polarized with electric field component $E_\omega$ within the π-plane spanned by $\mathbf{k}_i$ and $\mathbf{k}_f$, the wave-vectors of the incoming and outgoing beams (π-polarization). Two components of the scattered x-ray intensity are separated at the detector: $I_{\pi\pi}$, with $E_\omega$ remaining within the π-plane, and $I_{\pi\sigma}$, with $E_\omega$ now perpendicular to the π-plane. In the scattering geometry where $\mathbf{k}_i$, $\mathbf{k}_f \perp \mathbf{c}$, we have $I_{\pi\pi} \sim m^2_{//c}$ always (see Methods). We chose the incommensurate satellite reflections at (4+$q$, 4, 0) and (4, 4-$q$, 0) so that $I_{\pi\sigma} \sim m^2_{\perp q,c}$ and $I_{\pi\sigma} \sim m^2_{//q}$, respectively. Starting from the ZFC state at $T$ = 2.4 K and increasing the magnetic field, this convenient experimental configuration allows us to identify the helical ground state (H) with $m_{\perp q,c}$, $m_{//c} \neq 0$ and $m_{//q} = 0$ *(H=0,* Fig. 3**a,d,g**), the transverse conical (TC) state with $m_{//c} = 0$ and finite values for both in-plane components of **m(q)** ($\mu_0 H_{int}$ = 1.5 T, Fig. 3**b,e,h**), as well as the fan-like (F) state, which has only $m_{\perp q,c} \neq 0$ ($\mu_0 H_{int}$ = 2.9 T, Fig. 3**c,f,i**). It was confirmed that the incommensurate reflections vanish in the field-aligned state (not shown). The TC ground state in finite field is likely stabilized by weak in-plane anisotropy of the local magnetic moment. Weak in-plane anisotropy was also observed in magnetization measurements (Supplementary Figure 7).

In Methods (Fig. 5) and Supplementary Notes 2, 3 we present bulk neutron scattering data obtained on a $^{160}$Gd isotope-enriched single crystal. Firstly, we find excellent quantitative agreement of small-angle neutron scattering (SANS) and REXS, indicating that the REXS experiment is not seriously affected by surface strain and can be used to characterize the bulk properties of $Gd_3Ru_4Al_{12}$. Secondly, our neutron experiment with **H // a\*** confirms the multi-domain nature of the zero-field helical ground state. Thirdly, neutron scattering also provides proof that the magnetic modulations on the breathing kagomé layers are ferromagnetically stacked along the *c*-axis, by ruling out magnetic reflections at (*q*, 0, (2*n*-1)/2) and (*q*, 0, 2*n*-1) for *n*=1 and 2. As compared to the triangular lattice, kagomé structures introduce additional complexity due to the larger number of atoms per crystallographic unit cell. While the scope of this work does not include a full refinement of the magnetic structure, the interesting question of the local spin alignment on the trimer plaquette remains to be resolved in future studies.

We have also performed REXS experiments in the SkL phase at $T = 7$ K, $\mu_0 H_{int} = 1.5$ T under field cooling (Figs. 3**j-l**). In this experiment, the three reflections $(7+q, 0, 0)$, $(7, q, 0)$, and $(7+q, -q, 0)$ – corresponding to $\mathbf{q}_1$, $\mathbf{q}_2$, and $\mathbf{q}_3$ in the inset sketch of Fig. 3**l** - were chosen. We find very strong $I_{\pi\sigma}(\mathbf{q}_3) \sim m^2_{\perp q,c}$ but weaker $I_{\pi\sigma}(\mathbf{q}_1)$ and $I_{\pi\sigma}(\mathbf{q}_2)$, a telltale sign of the fan-like state. The large fan-like signal in this experiment likely arises due to the proximity to a first order phase transition and associated phase separation. Crucially, there is also significant $I_{\pi\pi} \sim m^2_{//c}$ with comparable intensities for all the three $\mathbf{q}_i$. Our data, taken with an x-ray beam spot size of ~1 mm$^2$, suggest about 20-50 % volume fraction ($f_V$) of the helical component $m_{//c}$. Roughly equal population of the scattering signal related to helical order for the three $\mathbf{q}_i$ is consistent with a topological multi-**q** ordered state, such as a lattice of Bloch skyrmions, in the SkL phase.

**Spin-vortices in real space imaging.** In the scattering study presented here, all information about the relative phase of the three helical modulations making up the SkL is lost. Thus, these experiments cannot confirm the topological nature of the SkL state. Meanwhile, imaging of the real-space spin structure using Lorentz transmission electron microscopy (L-TEM) on a thin-plate sample [29] provides unambiguous evidence for spin-vortices in the SkL state, as shown in the following (Fig. 4). For skyrmions with Bloch-type character, clockwise or counter-clockwise helicity (in-plane rotation direction of spins) should appear as bright and dark dots in the underfocused L-TEM images, respectively [3]. On the other hand, the helical structure with in-plane **q**-vector can be visualized as alternating bright and dark stripes in the underfocused L-TEM image.

All our L-TEM data was recorded under field cooled conditions. At $T = 8$ K and $\mu_0 H_{int} = 0.59$ T, we found stripe-like magnetic contrast of a single-**q** helical order with pitch $\lambda = 2.8$ nm (Fig. 4**a**-**b**). At the same temperature, but at higher field $\mu_0 H_{int} = 1.53$ T, the vortex-like pattern of our real space image translates into six-fold symmetric incommensurate reflections in the Fourier transform (Figs. 4**c**,**d**). In $T$-dependent measurements the magnetic contrast vanishes at $T = 17\text{-}20$ K (Fig. 4**e**-**h**). The region of the $B$-$T$ phase diagram occupied by the SkL phase was found to be slightly expanded in L-TEM measurements on thin plates of $Gd_3Ru_4Al_{12}$, as compared to bulk samples. This behavior is consistent with previous work on other skyrmion hosting materials [3,4]. Although the values of $\lambda$ from real space imaging and scattering experiments are in quantitative agreement, local lattice strain appears to rotate **q** towards the *a*-axis in the thin plate sample (see Methods for a discussion). The strain effect is also manifested in the fast Fourier transform data of our highest quality real space image, which is slightly distorted (area B, Fig. 4**d**).

With the aim of amplifying the weak contrasts of magnetic skyrmions and helical stripes in the L-TEM data, we cut background noise by preserving selected fast Fourier transform components as exemplified in Fig. 4**b**,**d**: (i) yellow circles mark in-plane **q**-vectors related to the helical structure and the SkL, while (ii) red circles mark the Fourier components related to the atomic crystal lattice of $Gd_3Ru_4Al_{12}$. The filtered fast Fourier transforms are then converted back to filtered real-space images as shown in the insets of Fig. 4**a,c**. Note that (ii) are visible only in the case of the high-field data measured at very small defocusing length of the electron beam (c.f. Methods, where more experimental details are provided). In combination with the scattering and transport experiments, our L-TEM study firmly establishes the presence of the SkL phase in this compound.

**Estimate of spin polarization in the conduction band from topological Hall effect.** Armed with microscopic knowledge of the magnetic order in the SkL phase, we proceed towards a semi-quantitative analysis of the topological Hall effect. In the continuum approximation, the emergent magnetic field from hexagonal lattice skyrmion textures is $B_{em} = -(h/e)\sqrt{3}/(2\lambda^2) \approx -460$ T [27,30]. This enormous effective field is related to the topological Hall resistivity through the normal Hall coefficient $R_0$, the volume fraction of the skyrmion phase $f_V$, and the effective spin polarization $P$ of conduction electrons [30]

$$\rho_{yx}^{THE} = f_V \cdot P \cdot R_0 \cdot B_{em} \qquad (2)$$

After extraction of the topological Hall conductivity $\sigma_{xy}^{THE}$ (Fig. 2**b**), we estimate the topological Hall resistivity as $\rho_{yx}^{THE} = \sigma_{xy}^{THE} \cdot \rho_{xx}^2$. Extrapolating the value of $R_0$ from higher temperatures (see Supplementary Note 5) and using $f_V = 20$-$100$ % in the SkL phase, we arrive at $P = 0.01$-$0.05$ (higher $f_V$ corresponds to lower $P$). These values of $P$ appear reasonable in comparison with related materials such as $Gd_2PdSi_3$ ($P = 0.07$) [19]. Note that the continuum approximation underlying Eq. (2) may be rendered inaccurate when $\lambda$ becomes comparable to the crystallographic lattice spacing. Hence, the observed magnitude of the THE should be taken as a merely semi-quantitative measure of $B_{em}$.

**Discussion.**

Our combined experimental effort shows that a topological SkL phase is stabilized in the centrosymmetric breathing kagomé lattice $Gd_3Ru_4Al_{12}$. This system charms with conceptual simplicity: Large local spins, whose moment size is affected little by thermal fluctuations, are coupled weakly to a sea of conduction electrons. In the SkL phase, the rare earth lattice imparts its scalar spin chirality onto the conduction electrons and drives a giant topological Hall effect in a limited window of temperature and magnetic field.

Superficially, the magnetic phase diagram (Fig. 2) suggests similarities with Bloch skyrmions in chiral magnets; the SkL is thermodynamically stable only at elevated temperatures, and a metastable skyrmion state survives at low $T$ under field-cooled conditions. Unlike in chiral magnets however, thermal fluctuations are by no means necessary to stabilize equilibrium topological spin textures in centrosymmetric lattices. On one hand, an equilibrium SkL was observed experimentally in triangular lattice $Gd_2PdSi_3$ down to very low temperature (at least $T/T_N=0.1$) [19]. On the other hand, numerical simulations have consistently shown extended parameter regimes with a ground state equilibrium SkL, both in the case of frustrated exchange interactions [31] and RKKY systems with four-spin-two-sites (biquadratic) couplings [32,33]. In the present system, we have mapped the RKKY interaction to an effective spin model which shows frustrated antiferromagnetic couplings at the second nearest neighbor level (Supplementary Note 4). As this is a metallic system however, treatment as a Kondo lattice with RKKY couplings and possibly sizeable four-spin (biquadratic) couplings may be more appropriate for a quantitative description. A prominent role for thermal fluctuations is re-introduced in the case of $Gd_3Ru_4Al_{12}$ due to a particularly delicate balance of terms in the free energy when the magnetic field is finite: Rather weak in-plane anisotropy (Supplementary Note 7) of the magnetic moments favors transverse conical (TC) order at the lowest $T$, while thermal fluctuations promote the SkL state. In consequence, metastable skyrmions could be observed in transport experiments for the first time in a centrosymmetric magnet. We stress that whereas Gaussian fluctuations of the moment amplitude provide the route to the SkL in non-centrosymmetric magnets [1], thermal fluctuations of the moment direction away from the easy plane of local anisotropy are expected to be essential in $Gd_3Ru_4Al_{12}$.

Our real space imaging study provides compelling evidence for incommensurate, topological multi-**q** order, for the first time in a centrosymmetric crystal with tiny (a few nanometer sized) magnetic vortices. More broadly, our results show that the breathing kagomé lattice is a rich host for new topological magnetic phases, which may exert a colossal emergent magnetic field to produce unprecedented electrical, thermal, and thermoelectric phenomena, like the large THE reported here.

**Methods.**

**Synthesis and characterization.** Large, cm-sized crystals of $Gd_3Ru_4Al_{12}$ were grown under Ar gas flow in a floating zone (FZ) furnace equipped for high vacuum operation. The samples were characterized by powder x-ray diffraction (XRD), energy-dispersive x-ray spectroscopy (EDX), and examination under an optical microscope equipped for polarization analysis. Laue x-ray diffraction was used to obtain samples with oriented surfaces. We also grew a crystal rod using the Czochralsky pulling technique, but found that the FZ approach is more stable and reproducibly yields high quality crystals. As Gd in natural abundance is a strong neutron absorber, an additional $^{160}$Gd isotope enriched single crystal batch was grown for elastic neutron scattering experiments. Due to oxide impurities in the $^{160}$Gd raw material, this growth was of lower quality as characterized by transport (RRR = $\rho(300\ K)/\rho(2\ K)$ ~ 2.6) and the sharpness of bulk phase transitions in the magnetization data. The final mass of the sample used for elastic neutron scattering was about 15 mg.

**Bulk measurement techniques.** Magnetization and heat capacity were measured using cube- or rectangular cuboid-shaped polished samples in commercial Quantum Design MPMS and PPMS cryostats, in an attempt to minimize adverse effects due to the demagnetization field. The magnetic field **H** was applied parallel to the crystallographic *c*-axis for all data shown in the main text. Due to experimental constraints, high-field vibrating sample magnetization (VSM) measurements were complemented by low-field ($\mu_0 H < 7$ T) extraction magnetization data (DC-M) on the same sample in the same configuration, to arrive at a more reliable estimate for the absolute value of *M*. The VSM data was then scaled to the DC-M results (scaling factor ~1.06). These measurements were performed in a Quantum Design PPMS-14T (VSM) and a Quantum Design MPMS3 (DC-M) system, respectively.

**Transport experiments.** Transport measurements were carried out using polished and oriented plates of approximate dimensions 2.5 x 1 x 0.15 mm$^3$ with electrical contacts attached by silver paste (Dupont) or H20E silver epoxy (Epo-Tek). For the data presented in Fig. 2, the face of the sample plate was perpendicular to the *c*-axis (**H** // **c**) and the charge current density was **J** // **a**\*. Similar data were obtained for a sample with **J** // **a** (Supplementary Figure 6). For the transport experiments, we used a Quantum Design PPMS cryostat for temperature and magnetic field control, but a custom arrangement of lockin- and pre-amplifiers replaced the PPMS measurement circuits. The applied current density was $J \sim 3 \cdot 10^4$ A/m$^2$ and the excitation frequency was 9-15 Hz. We calibrated the absolute value of $\rho_{xx}$ using a long, bar-shaped crystal of dimensions 3 x 0.5 x 0.5 mm$^3$ in order to reduce systematic errors arising from the measurement of the sample geometry when using a standard optical microscope. For **J** // **a** or **a**\*, the resistivity was 110(5) μΩcm at room temperature. The field dependence of $\rho_{xy}$ and $\rho_{xx}$ was calculated from anti-symmetrized and symmetrized voltage traces, respectively. Due to significant hysteresis at low temperature, we recorded full field ramps for both increasing and decreasing magnetic field and paired, for example, the data with d$H$/d$t$ < 0, $H$ < 0 with d$H$/d$t$ > 0, $H$ > 0 for the (anti-)symmetrization routine.

**Resonant elastic x-ray scattering (REXS).** Magnetic x-ray scattering experiments were conducted in resonance with the Gd $L_2$-edge on beamline BL-3A at Photon Factory in KEK, Japan with the scattering plane being ($H$, $K$, 0), i.e. the incoming and outgoing beams of wave vector $\mathbf{k}_i$ and $\mathbf{k}_f$ were both in the plane perpendicular to the crystallographic $c$-axis. The 006 reflection of a pyrolytic graphite (PG) plate was used to analyze the polarization of the scattered beam, with the $2\theta$ angle at the Gd-$L_2$ edge of PG fixed to 88 degrees. Our single crystal with large (110) planes was set in a cryostat equipped with a vertical 8 Tesla superconducting magnet, so that the magnetic field was applied parallel to the $c$-axis.

The expression for the magnetic part of the scattering amplitude in REXS is written as [34]

$$f_{res} = C_0 \boldsymbol{\varepsilon}_f^* \cdot \boldsymbol{\varepsilon}_i + iC_1(\boldsymbol{\varepsilon}_f^* \times \boldsymbol{\varepsilon}_i) \cdot \mathbf{m} + C_2 \boldsymbol{\varepsilon}_f^\dagger O \boldsymbol{\varepsilon}_i \qquad (3)$$

with constants $C_i$, local magnetization $\mathbf{m}$, and $\boldsymbol{\varepsilon}_i$ and $\boldsymbol{\varepsilon}_f$ the initial and final polarization of the x-ray beam, respectively. The scattering intensity is $I_{res} = f_{res} \cdot f_{res}^*$. Only the second term depends explicitly on $\mathbf{m}$. In our experiment, the incident beam had the linearly polarized electric field component in the scattering plane (π-polarization). Recalling that $\boldsymbol{\varepsilon} \cdot \mathbf{k} = 0$ for light, let us separate two components of the scattered radiation: (1) $\boldsymbol{\varepsilon}_i$ and $\boldsymbol{\varepsilon}_f$ are both in the scattering plane π. As the angle between $\mathbf{k}_i$ and $\mathbf{k}_f$ is defined as $2\theta$, it follows that $I_{\pi\pi}$ ~ $(\sin(2\theta)\, m_z)^2$. (2) $\boldsymbol{\varepsilon}_i$ is in the π-plane, and $\boldsymbol{\varepsilon}_f$ is perpendicular to it. Their cross product is directly proportional to $\mathbf{k}_i$, and we write $I_{\pi\sigma}$ ~ $(\mathbf{k}_i \cdot \mathbf{m})^2$. The π-σ intensity thus depends very sensitively on the relative alignment of the incoming beam and the modulated magnetization **m** = **m**(**q**). A comparison of magnetic scattering at different reflections (Fig. 3) can be used to separate the two components ($m_{\perp q,c}$, $m_{//q}$) of the in-plane magnetic moment.

**Lorentz transmission electron microscopy (L-TEM).** A (001) plate of $Gd_3Ru_4Al_{12}$ was cut from a bulk sample with dimensions 2 x 2 x 0.1 mm$^3$. The plate was further thinned by an Ar-ion milling process subsequent to mechanical grinding. Real-space observations were performed by L-TEM using a multifunctional transmission electron microscope (JEM2800, JEOL) equipped with a double-tilt helium cooling holder (ULTDT). We measured the transmittance of the incident electron beam with accelerating voltage of 200 kV. The temperature of the thin plate was carefully controlled from 7 K to 50 K.

L-TEM is a useful method to observe magnetic order in thin plates with thickness below several hundred nanometers due to the interactions between the incident electron beam and magnetic moments [29]. Data in low field ($\mu_0 H_{int}$ = 0.59 T, $\mu_0 H$ = 0.70 T, Fig. 4**a**,**b**) were obtained at defocusing length $l_d$= -3 μm, where only magnetic contrast is visible. The instrument cannot be operated at these same settings in high magnetic field ($\mu_0 H_{int}$ = 1.53 T corresponding to $\mu_0 H$ = 1.95 T, Fig. 4**c**-**h**): Instead, measurements were carried out at very small $l_d$ =-193 nm, so that both the crystal lattice and the magnetic order leave their fingerprints in the data. These high-field experiments conveniently show that the magnetic **q**-vector is aligned with the crystallographic *a*-axis in the thin plate sample, in contrast to the x-ray and neutron scattering results on bulk samples, where **q** // **a***. We found that further analysis of the data using the transport-of-intensity equation (TIE) approach was unreliable in the present case, where the size of the spin vortices is comparable to the real-space resolution of the Lorentz-TEM experiment and the magnetic contrast is relatively weak.

**Small angle neutron scattering (SANS) experiments.** For our neutron scattering study, we used the time-of-flight (TOF) type small-and-wide-angle neutron scattering instrument TAIKAN(BL15) at the Material and Life Science Experimental Facility (MLF) of J-PARC in Japan. Pulsed neutrons from a spallation source with a distribution of wavelengths centered around $\lambda_{n,max}$ = 2.5-3 Å are incident on the sample; some of the neutrons are scattered elastically from the crystal and deflected so that $\mathbf{k}_f \neq \mathbf{k}_i$ ($\mathbf{k}_i$, $\mathbf{k}_f$ being the wave vector of the incoming and outgoing neutron, respectively). The time delay between the spallation event and the arrival at the detector bank allows to extract the wavelength $\lambda_n = 2\pi k_{in}^{-1}$ of each neutron, while its position on the detector screen corresponds to the scattering angle $2\theta$ between $\mathbf{k}_i$ and $\mathbf{k}_f$. From $\lambda_n$ and $2\theta$, the components $Q_x$, $Q_y$, and $Q_z$ of the scattering vector $\mathbf{Q} = \mathbf{k}_i - \mathbf{k}_f$ can be calculated.

Fig. 5**a** shows a two-dimensional $Q_x$-$Q_y$ map of scattering intensity (a $Q_z = 0$ cut of the full TOF data set) measured using the small-angle neutron scattering (SANS) detector bank of TAIKAN in zero field and $T$ = 3.58 K. This data, obtained before the resonant x-ray scattering experiment discussed in the main text, provided the first evidence for incommensurate magnetic order in $Gd_3Ru_4Al_{12}$. The magnetic modulation vector was found to be along the **a**\* crystallographic direction and those other directions equivalent to **a**\* by hexagonal symmetry. The six-fold symmetric pattern on the detector arises due to scattering from multiple domains with different alignment of **q**, as we show in Supplementary Note 2. Due to the relatively large absolute value of $q = 0.272$ r.l.u. as compared to, for example, MnSi [2], our sample has to be tilted slightly away from $\mathbf{k}_i$ // **c** in order to observe the SANS reflections corresponding to incommensurate magnetic order (i.e., if $\mathbf{k}_i$ // **c**, **Q** is not perfectly aligned within the *a-b* plane). In such a tilted configuration, only the three reflections with $Q_x > 0$ are accessible out of the total six peaks allowed by hexagonal symmetry. However, an additional reflection at $Q_x < 0$ in Fig. 5**a** arises due to a second crystallographic domain in the $^{160}$Gd enriched sample. We estimate that the minority domain has about 25% volume fraction of the total crystal, with its *c*-axis about 6° misaligned from the *c*-axis of the major domain.

After submission of this manuscript we became aware of a recent resonant x-ray scattering study on the helical, zero-field ground state of $Gd_3Ru_4Al_{12}$ [35].

# References.

1. Roessler, U.K., Bogdanov, A.N. & Pfleiderer, C. Spontaneous skyrmion ground states in magnetic metals. Nature **442**, 797-801 (2006)

2. Mühlbauer, S., Binz, B., Jonietz, F., Pfleiderer, C., Rosch, A., Neubauer, A., Georgii, R. & Böni, P. Skyrmion Lattice in a Chiral Magnet. Science **323**, 915-919 (2009)

3. Yu, X.Z., Onose, Y., Kanazawa, N., Park, J.H., Han, J.H., Matsui, Y., Nagaosa, N. & Tokura, Y. Real-space observation of a two-dimensional skyrmion crystal. Nature **456**, 901-904 (2010)

4. Nagaosa, N. & Tokura, Y. Topological properties and dynamics of magnetic skyrmions. Nature Nanotechnology **8**, 899-911 (2013)

5. Schulz, T., Ritz, R., Bauer, A., Halder, M., Wagner, M., Franz, C., Pfleiderer, C., Everschor, K., Garst, K. & Rosch, A. Emergent electrodynamics of skyrmions in a chiral magnet. Nature Physics **8**, 301-304 (2012)

6. Garel, T. & Doniach, S. Phase transitions with spontaneous modulation-the dipolar Ising ferromagnet. Physical Review B **26**, 325 (1982)

7. Kézsmárki, I., Bordács, S., Milde, P., Neuber, E., Eng, L.M., White, J.S., Rønnow, H.M., Dewhurst, C.D., Mochizuki, M., Yanai, K., Nakamura, H., Ehlers, D., Tsurkan, V. & Loidl, A. Néel-type skyrmion lattice with confined orientation in the polar magnetic semiconductor $GaV_4S_8$. Nature Materials **14**, 1116 (2015)

8. Matsuno, J., Ogawa, N., Yasuda, K., Kagawa, F., Koshibae, W., Nagaosa, N., Tokura, Y. & Kawasaki, M. Interface-driven topological Hall effect in $SrRuO_3$-$SrIrO_3$ bilayer. Science Advances **2**, e1600304 (2016)

9. Wang, L., Feng, Q., Kim, Y., Kim, R., Lee, K.H., Pollard, S.D., Shin, Y.J., Zhou, H., Peng, W., Lee, D., Meng, W., Yang, H., Han, J.H., Kim, M., Lu, Q. & Noh, T.W. Ferroelectrically tunable magnetic skyrmions in ultrathin oxide heterostructures. Nature Materials **17**, 1087 (2018)

10. Okubo, T., Chung, S., & Kawamura, H. Multiple-$q$ States and the Skyrmion Lattice of the Triangular-Lattice Heisenberg Antiferromagnet under Magnetic Fields. Physical Review Letters **108**, 017206 (2012)

11. Leonov, A.O. & Mostovoy, M. Multiply periodic states and isolated skyrmions in an anisotropic frustrated magnet. Nature Communications **6**, 8275 (2015)

12. Gao, S., Zaharko, O., Tsurkan, V., Su, Y., White, J.S., Tucker, G.S., Roessli, B., Bourdarot, F., Sibille, R., Chernyshov, D., Fennell, T., Loidl, A. & Rüegg, C. Spiral spin-liquid and the emergence of a vortex-like state in $MnSc_2S_4$. Nature Physics **13**, 157 (2017)

13. Heinze, S., von Bergmann, K., Menzel, M., Brede, J., Kubetzka, A., Wiesendanger, R., Bihlmayer, G. & Blügel, S. Spontaneous atomic-scale magnetic skyrmion lattice in two dimensions. Nature Physics **7**, 713 (2011)

14. Hayami, S. & Motome, Y. Multiple-$Q$ instability by ($d-2$)-dimensional connections of Fermi surfaces. Physical Review B **90**, 060402(R) (2014)

15. Takagi, R., White, J.S., Hayami, S., Arita, R., Honecker, D., Rønnow, H.M., Tokura, Y. & Seki, S. Multiple-$q$ noncollinear magnetism in an itinerant hexagonal magnet. Science Advances **4**, eaau3402 (2018)

16. Lin, S.-Z. & Hayami, S. Ginzburg-Landau theory for skyrmions in inversion-symmetric magnets with competing interactions. Physical Review B **93**, 064430 (2016)

17. Tokura, Y. & Nagaosa, N. Nonreciprocal responses from non-centrosymmetric quantum materials. Nature Communications **9**:3740 (2018)

18. Desplat, L., Kim, J.-V. & Stamps, R.L. Paths to annihilation of first- and second-order (anti)skyrmions via (anti)meron nucleation on the frustrated square lattice. Physical Review B **99**, 174409 (2019)

19. Kurumaji, T., Nakajima, T., Hirschberger, M., Kikkawa, A. Yamasaki, Y., Sagayama, H., Nakao, H., Taguchi,Y., Arima, T. & Tokura, Y. Skyrmion lattice with a giant topological Hall effect in a frustrated triangular-lattice magnet. Science **365**, 914-918 (2019)

20. Gladyshevskii, R. E., Strusievicz, O. R., Cenzual, K. & Parthé, E. Structure of $Gd_3Ru_4Al_{12}$, a new member of the $EuMg_{5.2}$ structure family with minority-atom clusters. Acta Crystallographica Section B **49**, 474-478 (1993)

21. Niermann, J. & Jeitschko, W. Ternary Rare Earth (*R*) Transition Metal Aluminides $R_3T_4Al_{12}$ (*T* = Ru and Os) with $Gd_3Ru_4Al_{12}$ Type Structure. Journal for Inorganic and General Chemistry, **628**, 2549-2556 (2002)

22. Nakamura, S., Kabeya, N., Kobayashi, M., Araki, K., Katoh, K. & Ochiai, A. Spin trimer formation in the metallic compound $Gd_3Ru_4Al_{12}$ with a distorted kagome lattice structure. Physical Review B **98**, 054410 (2018)

23. Chandragiri, V., Iyer, K. K. & Sampathkumaran, E.V. Magnetic behavior of $Gd_3Ru_4Al_{12}$, a layered compound with distorted kagomé net. Journal of Physics: Condensed Matter **28**, 286002 (2016)

24. Sato, T., Kadowaki, H., Masuda, H. & Iio, K. Neutron Diffraction Study of Successive Phase Transitions in the Heisenberg Antiferromagnet $MnBr_2$. Journal of the Physical Society of Japan **63**, 4583 (1994)

25. Kenzelmann, M., Harris, A. B., Jonas, S., Broholm, C., Schefer, J., Kim, S.B., Zhang, C.L., Cheong, S.-W., Vajk, O. P. & Lynn, J.W. Magnetic Inversion Symmetry Breaking and Ferroelectricity in $TbMnO_3$. Physical Review Letters **95**, 087206 (2005)

26. Inami, T., Terada, N., Kitazawa, H. & Sakai, O. Resonant Magnetic X-ray Diffraction Study on the Triangular Lattice Antiferromagnet $GdPd_2Al_3$. Journal of the Physical Society of Japan **78**, 084713 (2009)

27. Neubauer, A., Pfleiderer, C., Binz, B., Rosch, A., Ritz, R., Niklowitz, P. G. & Böni, P. Topological Hall Effect in the A Phase of MnSi. Physical Review Letters **102**, 186602 (2009)

28. Lee, M., Kang, W., Onose, Y., Tokura, Y. & Ong, N. P. Unusual Hall Effect Anomaly in MnSi under Pressure. Physical Review Letters **102**, 186601 (2009)

29. Reimer, L. & Kohl, H. Transmission Electron Microscopy. 5th Edition, P267-P271 Springer (1997)

30. Ritz, R., Halder, M., Franz, C., Bauer, A., Wagner, M., Bamler, R., Rosch, A. & Pfleiderer, C. Giant generic topological Hall resistivity of MnSi under pressure. Physical Review B **87**, 134424 (2013)

31. Hayami, S., Lin, S.-Z. & Batista, C.D. Bubble and skyrmion crystals in frustrated magnets with easy-axis anisotropy. Physical Review B **93**, 184413 (2016)

32. Ozawa, R., Hayami, S. & Motome, Y. Zero-Field Skyrmions with a High Topological Number in Itinerant Magnets. Physical Review Letters **118**, 147205 (2017)

33. Hayami, S., Ozawa, R. & Motome, Y. Effective bilinear-biquadratic model for noncoplanar ordering in itinerant magnets. Physical Review B **95**, 224424 (2017)

34. Lovesey S. W. & Collins, S.P. X-ray Scattering and Absorption by Magnetic Materials. Clarendon Press (1996)

35. Matsumura, T., Ozono, Y., Nakamura, S., Kabeya, N. & Ochiai, A. Helical Ordering of Spin Trimers in a Distorted Kagome Lattice of $Gd_3Ru_4Al_{12}$ Studied by Resonant X-ray Diffraction. Journal of the Physical Society of Japan **88**, 023704 (2019)

**Acknowledgements.**

We are grateful for helpful discussions with Naoto Nagaosa, Shinichiro Seki, and Nguyen Khanh. Yukitoshi Motome and Satoru Hayami provided stimulating comments on an early draft of this manuscript, which we thankfully acknowledge. We are also indebted to Kiyomi Nakajima for technical assistance and to Daisuke Hashizume for making his in-house x-ray spectrometer available to us for structural analysis. Resonant x-ray scattering measurements were performed under the approval of the Photon Factory program advisory committee (Proposals No. 2015S2-007) at the Institute of Material Structure Science, High Energy Accelerator Research Organization (KEK). Neutron scattering measurements at the Materials and Life Science Experimental Facility of J-PARC were performed under a user program (Proposal No. 2017L0700). M. H. was supported as a JSPS International Research Fellow (18F18804). This project was also supported by Grant-In-Aid for Young Scientists(B) No. 17K14351 and by JST CREST Grant Number JPMJCR1874 (Japan).

**Author Contributions:**

M.H. and A.K. grew and characterized the single crystals with initial help from T.K. Transport, magnetization, and heat capacity measurements were carried out and analyzed by M. H. with support from M.K. T.N. and M.H. measured the neutron scattering data with technical assistance from K.O. T.N. and S.G. carried out the x-ray scattering experiments supported by Y.Y., H.S., and H.N. L.P. and X.Y. prepared and measured the Lorentz-TEM sample. M.H. and Y.To. wrote the manuscript with contributions from all authors. Y. To., T.A., K.K., and Y. Ta. designed and oversaw the project.

**Competing interests**

The authors declare no competing interests.

**Data availability**

The data presented in the current study are available from the corresponding authors on reasonable request.

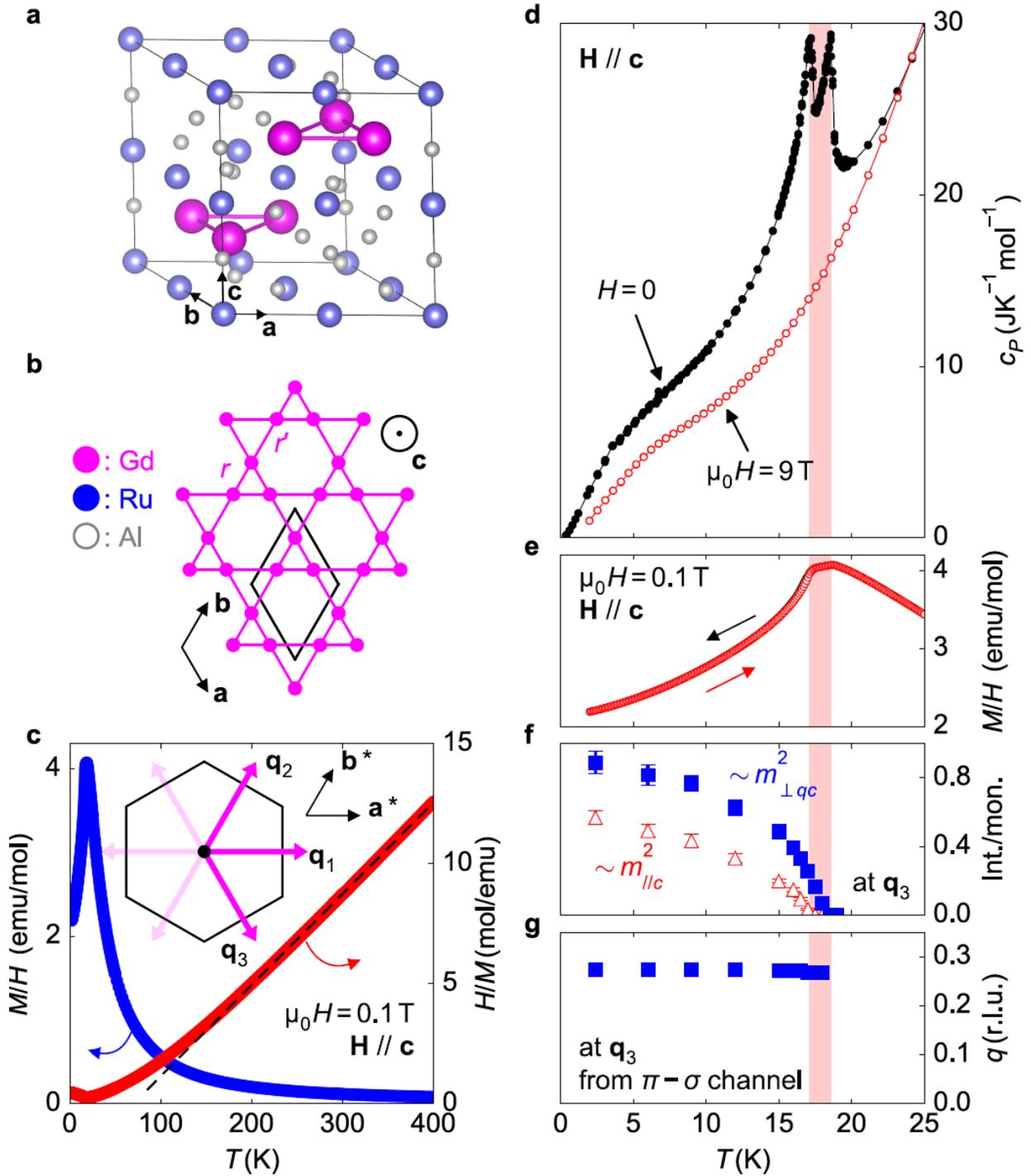

**Figure 1: Crystal structure and zero-field magnetic order of a Gd-based breathing kagomé lattice. a**, Hexagonal unit cell of $Gd_3Ru_4Al_{12}$, where *a*, *b*, and *c* are crystallographic lattice directions. **b**, Within the $Gd_3Al_4$ layer, rare earth (Gd) atoms form

a distorted kagomé net with alternating distances $r$, $r'$ between nearest neighbors. Al and Ru atoms are not shown. The black rhombus indicates the size of the primitive unit cell. **c**, Magnetic susceptibility (blue, left axis) increases continuously in the paramagnetic state as temperature is lowered. The inverse susceptibility $H/M$ (red, right axis) is fitted by the Curie-Weiss expression (dashed line) at high temperature. **d,e**, Specific heat $c_P(T)$ and $M/H$ show two phase transitions in zero magnetic field. **f,g**, At the $(7, 0, 0)+\mathbf{q}_3 = (7+q, -q, 0)$ incommensurate reflection, resonant x-ray scattering with polarization analysis provides modulated moments within ($m_{\perp q,c}$, blue solid triangles) and perpendicular to ($m_{//c}$, red open triangles) the hexagonal plane, as well as the magnitude of the ordering vector $\mathbf{q}$. Inset of **c**, six directions of $\mathbf{q}_i$ are allowed by symmetry. The black hexagon indicates a conventional unit cell in real space. The transition temperatures $T_{N2} > T_{N2}$ bound the red shaded area in **d**-**g**.

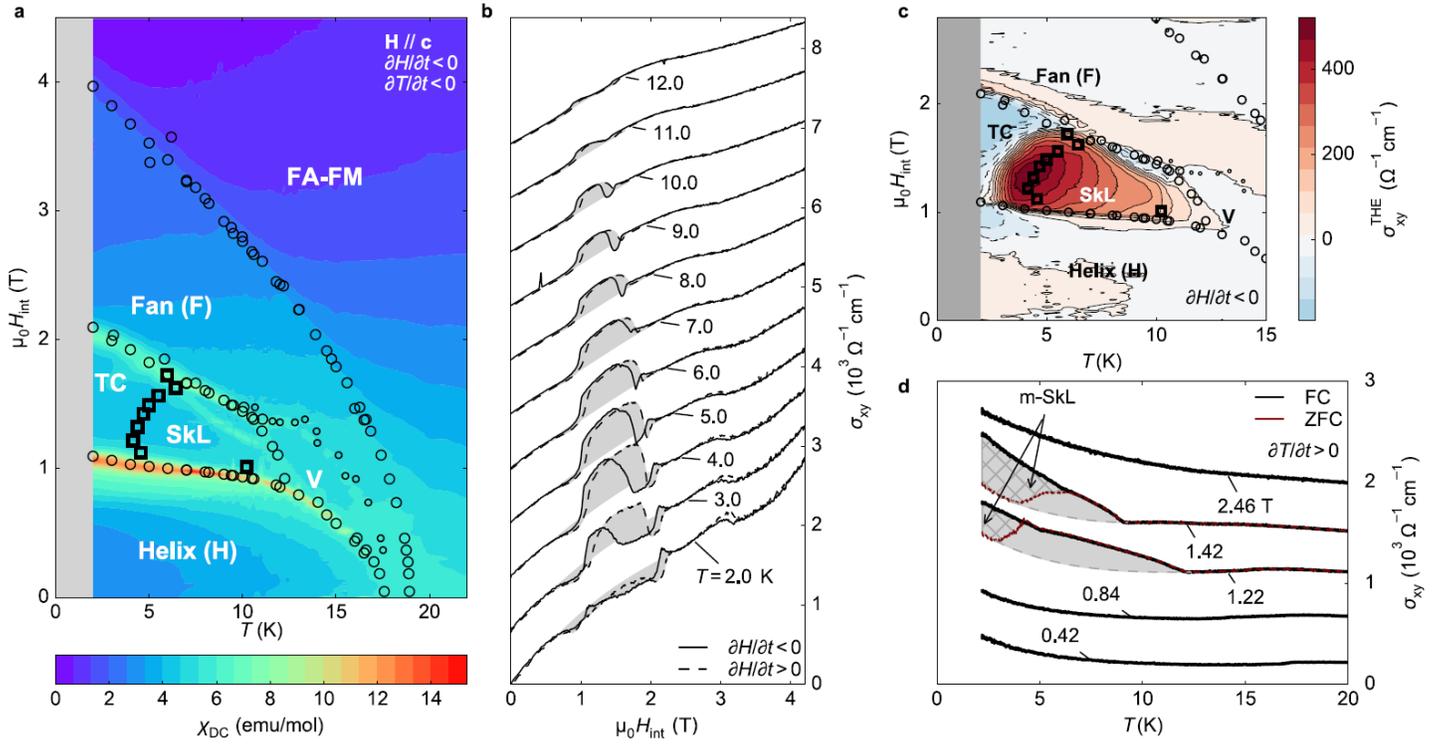

**Figure 2: Field-induced phase transitions in $Gd_3Ru_4Al_{12}$ and topological Hall effect in the SkL phase. a.** Contour plot of magnetic susceptibility $\chi_{DC} = \partial M/\partial H$ in which we distinguish helical (H), fan-like (F), transverse conical (TC), and skyrmion lattice (SkL) states, as well as the as-yet unidentified phase (V). FA-FM labels the field-aligned ferromagnetic state. Black open circles are phase boundaries obtained from $\chi_{DC}$ for $\partial H/\partial t$, $\partial T/\partial t <0$ (small symbols bound phase V. c.f. Supplementary Note 7). The transition between SkL and TC is marked by black open squares, corresponding to the onset of hysteresis between FC and zero-field cooled (ZFC) cases in panel **d**. **b**. Isotherms of the Hall conductivity $\sigma_{xy}$ at various temperatures. Dashed and solid lines mark increasing and decreasing magnetic fields, respectively. A polynomial background was subtracted to identify the topological Hall conductivity $\sigma_{xy}^{THE}$ (grey shaded area). Curves shifted by offsets of $\Delta\sigma_{xy} = 680$ $\Omega^{-1}$cm$^{-1}$. **c**, Large positive $\sigma_{xy}^{THE}$ obtained from **b** is observed only in the SkL phase. Black open dots and squares indicate phase boundaries as in **a**. **d**, $\sigma_{xy}(T)$ at fixed external magnetic field, for FC (bold, black line) and ZFC (thin, red line) conditions. Outside the TC phase, ZFC and FC curves overlap. In the TC phase, the FC signal is larger due to the additional $\sigma_{xy}^{THE}$ of metastable skyrmions (m-SkL, grey cross-hatched area). In the SkL state itself, we have estimated the background of $\sigma_{xy}$ with a polynomial (dark grey dashed line for $\mu_0 H_{int} = 1.22$ and $1.42$ T) and marked the $\sigma_{xy}^{THE}$ by grey coloring (no hatching). Curves shifted by offsets of $\Delta\sigma_{xy} = 250$ $\Omega^{-1}$cm$^{-1}$ for clarity.

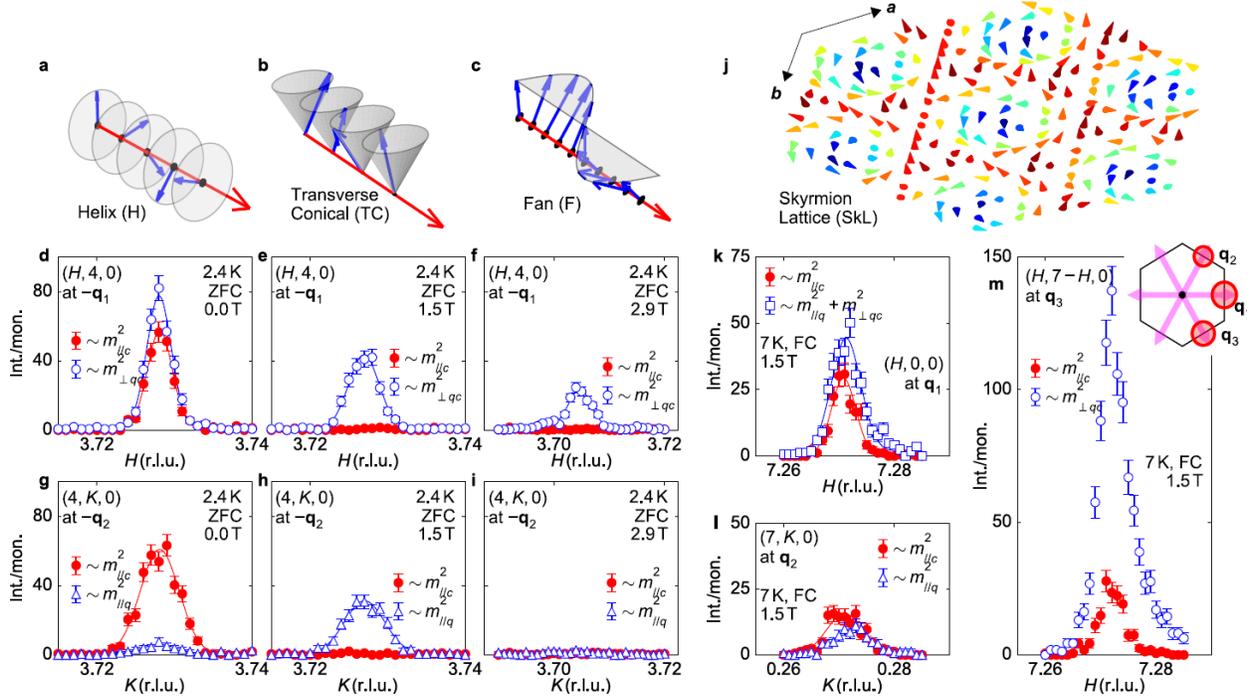

**Figure 3: Resonant elastic x-ray scattering (REXS) with polarization analysis clarifies the magnetic order of bulk Gd$_3$Ru$_4$Al$_{12}$.** The scattering plane is $(H, K, 0)$. Blue open symbols correspond to $I_{\pi\sigma}$ scattering intensity, related to magnetization components

in the hexagonal basal plane ($m^2_{//q}$, $m^2_{\perp q,c}$), while red solid symbols are for $I_{\pi\pi}$ intensities, proportional to $m^2_{//c}$. Data in **d-i** was obtained in the zero-field cooled (ZFC) state, whereas **j-l**, report field-cooled (FC) measurements. **a-i**, The comparison of scattering intensities at (4-$q$, 4, 0) and (4, 4-$q$, 0) reflections allows to separate the two in-plane components $m_{//q}$ and $m_{\perp q,c}$ for the three magnetic orders sketched in **a-c**, observed in the sequence (**a**-**b**-**c**) as the magnetic field is increased starting from the ZFC state at $T = 2.4$ K. Only the helical state in **d,g** has $m_{//c} \neq 0$. **k-m**, At $T = 7$ K and $\mu_0 H_{int} = 1.5$ T, the fan-like state coexists with helical modulations under FC conditions. As the fan-like state has $m_{//c} = 0$, $I_{\pi\pi}$ scattering directly probes the helical component. Comparable magnitude of $I_{\pi\pi} \sim m^2_{//c}$ at the three incommensurate reflections around (7, 0, 0) is consistent with multi-**q** helimagnetic order. The nano-scale skyrmion lattice (SkL) proposed here is sketched in **j**, with each cone representing the magnetic moment on an individual Gd-site. Red (blue) correspond to positive (negative) $m_z$ component of the local moment. Due to the lack of full refinement of the Gd-site moment, **j** visualizes but one possible manifestation of the SkL order. Inset of **m**, red circles of radius $r_i \sim \sqrt{I_{\pi\pi}} \sim m_{//c}$, illustrating approximately equal weight for the three directions **q**$_1$, **q**$_2$, and **q**$_3$ marked by light pink arrows. Error bars correspond to one standard deviation.

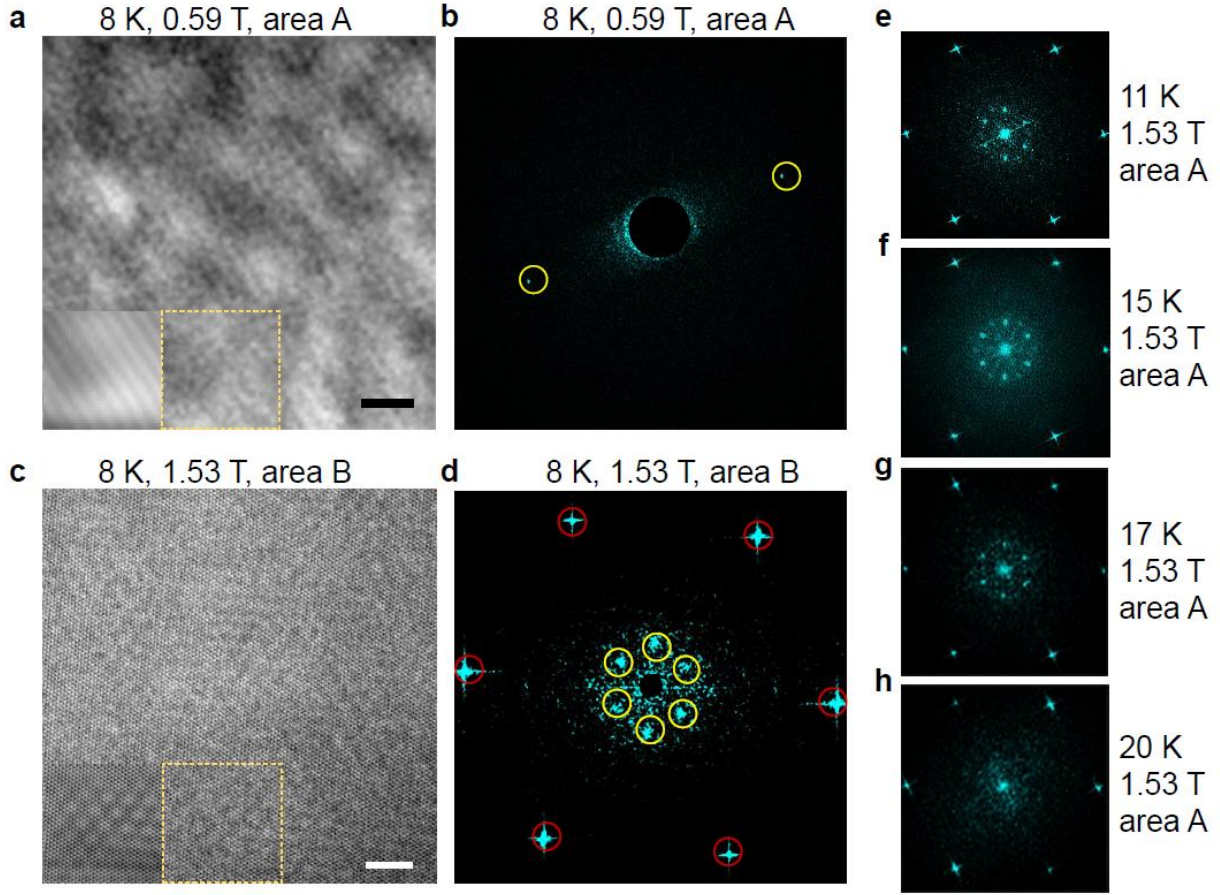

**Figure 4: Real-space observations of helical stripes and SkL in a (001) thin-plate sample of $Gd_3Ru_4Al_{12}$ using Lorentz-transmission electron microscopy (L-TEM).** We report data obtained from two different areas A, B of the same thin plate. The magnetic field was applied perpendicular to the sample plane. All data were recorded under field cooling (FC) conditions. **a*** and **b*** are crystalline axes in reciprocal space. Fast Fourier transform patterns in **b,d** correspond to real-space images (underfocused L-TEM images) shown in **a** and **c**, respectively. The defocusing length of the electron beam was $l_d$ = -3 μm (**a,b**) and $l_d$ = -193 nm (**c-h**). In **b**, the scale of the data was magnified as compared to **d,** cutting the featureless high-$q$ regime for clarity. Red (yellow) circles in **b,d** mark

intensity due to the crystal lattice (the magnetic order). Focusing on representative parts of the real space image (yellow dashed box), filtered images are shown as insets in **a**, **c** (same length scale as main panel). Helical stripes ($T = 8$ K, $\mu_0 H_{int} = 0.59$ T, $\mu_0 H = 0.7$ T) and SkL ($T = 8$ K, $\mu_0 H_{int} = 1.53$ T, $\mu_0 H = 1.95$ T) are visible in the data. **e-h**, $T$-dependence of the fast Fourier transform patterns of the magnetic and lattice images at 1.53 T. The magnetic contrast of the SkL is suppressed above $T = 17$ K. See Methods for filtering procedure and other experimental details. Scale bars in panels **a** and **c** correspond to 10 nanometers.

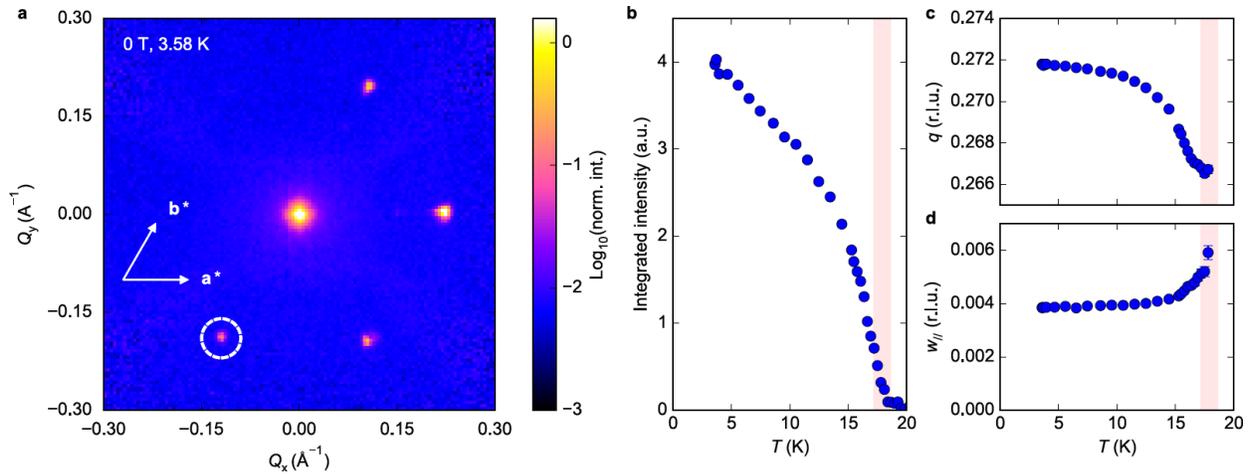

**Figure 5: Small angle neutron scattering (SANS) experiments as obtained at TAIKAN for $^{160}$Gd enriched Gd$_3$Ru$_4$Al$_{12}$ in H = 0 and at low $T$. a.** $Q_z = 0$ cut of scattering intensity in the SANS experiment. Three magnetic reflections corresponding to the multi-domain helical ground state are observed. A fourth reflection (marked by a white circle) originates from a minority crystallographic domain with different alignment of the $c$-axis. Scattered neutron intensity is shown on a logarithmic scale. **b.** Integrated intensity

of elastic neutron scattering onsets around $T_{N2}$ = 18.6 K. **c.** Weak temperature dependence of the absolute value of the magnetic modulation vector $q = 2\pi/\lambda$ on the order of 2% was observed. The values of $q$ obtained independently from REXS and SANS on different crystal batches are in good agreement. **d**, Half width at half maximum ($w_{//}$) of the magnetic reflection in elastic neutron scattering increases as $T_{N2}$ is approached from below. Red shaded area in **b**-**d** is bounded by $T_{N1}$, $T_{N2}$.